\documentclass[twoside,slac_one]{revtex4}
\usepackage{graphicx}
\usepackage{fancyhdr}
\usepackage{amsmath} 
\usepackage{bm}
\usepackage{amsxtra}
\usepackage{amssymb}
\usepackage{amsthm}
\usepackage{latexsym}
\usepackage{lscape}

\newcommand{\plots}{./}
\newcommand{\TeV}{\ensuremath{\mathrm{Te\kern -0.1em V}}}
\newcommand{\GeV}{\ensuremath{\mathrm{Ge\kern -0.1em V}}}
\newcommand{\gevcc}{\ensuremath{\mathrm{Ge\kern -0.1em V}/ c^{2}}}
\newcommand{\gevc}{\ensuremath{\mathrm{Ge\kern -0.1em V}/ c}}
\newcommand{\pt}{\ensuremath{p_{\rm T}}}
\newcommand{\met}  {\mbox{${\hbox{$E$\kern-0.6em\lower-.01ex\hbox{/}}}$}\ensuremath{_{\rm T}}}
\newcommand{\mpt}  {\mbox{${\hbox{$P$\kern-0.6em\lower-.01ex\hbox{/}}}$}\ensuremath{_{\rm T}}}
\newcommand{\et}{\ensuremath{E_{T}}}

\begin{document}

\title{Searches For the Standard Model Higgs boson in final states with $\tau$ leptons in $p\bar{p}$ collision at {\bf $\sqrt{s} = 1.96 \ \TeV$}}

%

\author{E. Pianori}
\affiliation{Department of Physics and Astronomy, University of Pennsylvania, Philadelphia, PA, USA}

\begin{abstract}
We present the status of the Standard Model Higgs boson searches in final states containing $\tau$ leptons using data collected
from $p\bar{p}$ collisions at Fermilab Tevatron collider at $\sqrt{s}= 1.96 \ \TeV$. A summary of the latest
results from CDF and D0 collaborations is reported in this paper.
\end{abstract}

\maketitle

\thispagestyle{fancy}


\section{Introduction}

The Higgs mechanism, proposed in the 1960 by P. Higgs to explain the origin of the electroweak symmetry breaking and particle masses \cite{Higgs},
predicts the presence of a yet to be discovered neutral scalar particle, called the Higgs boson. Since the Higgs boson couplings to the Standard Model 
particles are fixed by the particle masses,
the only free parameter of the theory is the Higgs boson mass. Indirect constrains from precision electroweak measurements prefer 
a Higgs boson with mass $m_{H} = 92^{+34}_{-26} \ \gevcc$
at 68\% confidence level, with a 95\% upper limit of $158 \ \gevcc$ \cite{EWKFit}. Direct limits set at LEP excluded $m_{H} < 114.4 \  \gevcc$  at 95\% C.L. level \cite{lephiggs}.\\
In the last few years CDF and D0 have steadily increased the sensitivity of the Higgs boson searches: as of July 2011, the combined
results from both the Tevatron experiments excluded at 95\% C.L. a Higgs boson in the range $100 \le m_{H} \le 108 \ \gevcc$ and $156 \le m_{H} \le 177 \ \gevcc$ \cite{combination}.\\
A constant effort has been made to improve analysis of the most sensitive channels, and an effort
have been started to include analysis of channels with lower expected sensitivity.
The searches that include $\tau$ leptons in the final state fall into this last category, and will be discussed in this paper.\\
There are two classes of analyses that include $\tau$ leptons that will be explored:
searches that target $H \to \tau\tau$ final states, and searches where the $W$ boson presents in the final state decays as $W\to\tau\nu$.\\
The former ones are important at low mass, where the BR($H\to \tau\tau$) is $\sim$ 7.4\% for $m_{H}=115 \ \gevcc$: not only they contribute to improve
the low mass sensitivity, but also, in case the Higgs boson exists, will allow to measure some of its properties. 
The latter one can be considered an addition to the main analyses, such as $WH \to l \nu b\bar{b}$ and $H\to WW$,
that target final states where the $W$ decays into $e$ and $\mu$.

\section{Experimental Apparatus}

The CDF and D0 detectors, already described in details in \cite{D0} and \cite{CDF}
are used to perform the searches presented in this paper.
The Tevatron, 
that ceased its operation on September 30th 2011,
provided $p\bar{p}$ collisions at $\sqrt{s} = 1.96 \ \TeV$. 
Since the beginning or Run II, over 12 $fb^{-1}$ of data were delivered at the two collision points, 
and almost $10 fb^{-1}$ of data were made available for the analysis to each experiment.
The results discussed here were performed using data corresponding to  $5.5 < \int L <  8.2 \ fb^{-1}$.

\section{Object reconstruction} \label{sec:tauID}

Tau leptons are short-lived particles which can be detected only through their visible decay
products. Hadronic decays, denoted as $\tau_{h}$, are of the form $\tau \to X_{h}\nu_{\tau}$ , where $X_{h}$ is a system of
hadrons, mainly composed of charged and neutral pions: the corresponding B.R. is about 65\%.
The remaining channels are represented by the leptonic decays
into muons ($\tau_{\mu}$ ) or electrons ($\tau_{e}$), through the process $\tau \to l \nu_{l} \nu_{\tau}$ . Since it is not possible to distinguish leptonically decaying $\tau$'s from $e$ and $\mu$
produced in the primary interaction, when talking about $\tau$'s throughout the paper we will mean
hadronically decaying ones.
The best quantity to describe the energy of the decay products of the 
$\tau$ is the visible four-momentum, defined as the sum of the track momenta and the neutral pions.
Because of its signature in the detector as a narrow calorimeter clusters matched to low
multiplicity reconstructed tracks, a $\tau$ can be easily mimicked by a jet.
Since $\tau$'s from $W$, $Z$ and $H$ decays are isolated and boosted, it is possible to exploit 
isolation-related observables to improve the purity of the $\tau$ identification algorithm.\\
The $H\to \tau\tau$+jets analysis of CDF and all the D0 analyses dicussed here use Multivariate Technique
based algorithm to identify the $\tau$ leptons.
The NN used by D0 is trained for three different decay modes of the $\tau$ separately: $\tau \to \pi^{\pm}\nu$ (type-I), $\tau \to \pi^{\pm}\pi^{0}\nu$ (type-II) 
and $\tau \to \pi^{\pm}\pi^{\pm}\pi^{\mp}(\pi{0})\nu$ (type-III). The CDF Boosted Decision Tree \cite{BDT} 
was developed independently for 1-prong and 3-prong $\tau$ candidates.
More details can be found in \cite{htt_cdf}, \cite{tauD0}.
All the other CDF analyses use cut based identification algorithms, optimized to the particular final 
state and trigger in use.\\
\indent Jets are reconstructed as clustered energy depositions in the calorimeter towers
 using a cone algorithm, with cone size $R = \sqrt{ (\Delta \Phi)^{2} + (\Delta \eta)^{2}} = 0.4(0.5)$
for CDF(D0).\\
\indent The missing transverse energy $\met$, calculated as the vector sum of all the calorimeter tower energy depositions projected to the transverse plane, is
used to determine the presence of neutrinos in the event. 

\section{Searches for $H \to \tau \tau$ final states}
For  Higgs boson masses $m_{H} \le 135 \ \gevcc$, the branching ratio BR($H\to \tau\tau$) is in the range of $4-8 \%$.
Depending on the production mechanism, other objects other than the $\tau$ candidates will be present in the events: the analyses discussed in this
review include extra jets or extra leptons in the final state.\\
In case the Higgs boson exists, it will be crucial to measure its couplings to different particles. 
The $\tau-\tau$ final states can be used to measure both the BR($H\to \tau\tau$) and the Higgs particle couplings to gauge bosons.
A significant deviation of one of these quantities from its expected value
could be evidence of new physics at play, since the branching ratios and 
production cross sections differ in Standard Model and Beyond the Standard Model scenarios.

\subsection{$H \to \tau \tau$+ jets}

\begin{figure}[h]
\begin{center}
\includegraphics[width=0.31\textwidth,height=0.15\textheight]{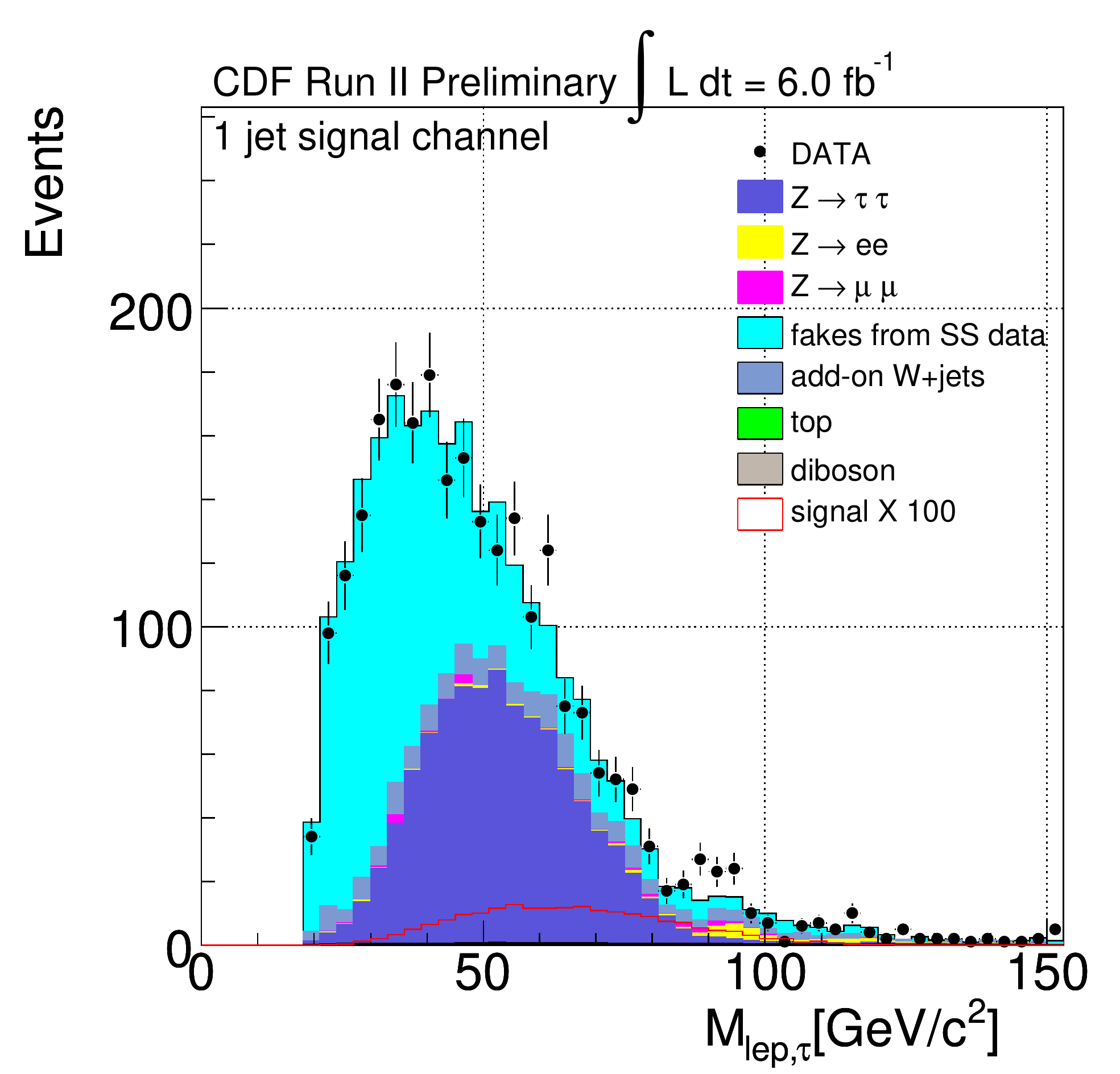}
\includegraphics[width=0.31\textwidth,height=0.15\textheight]{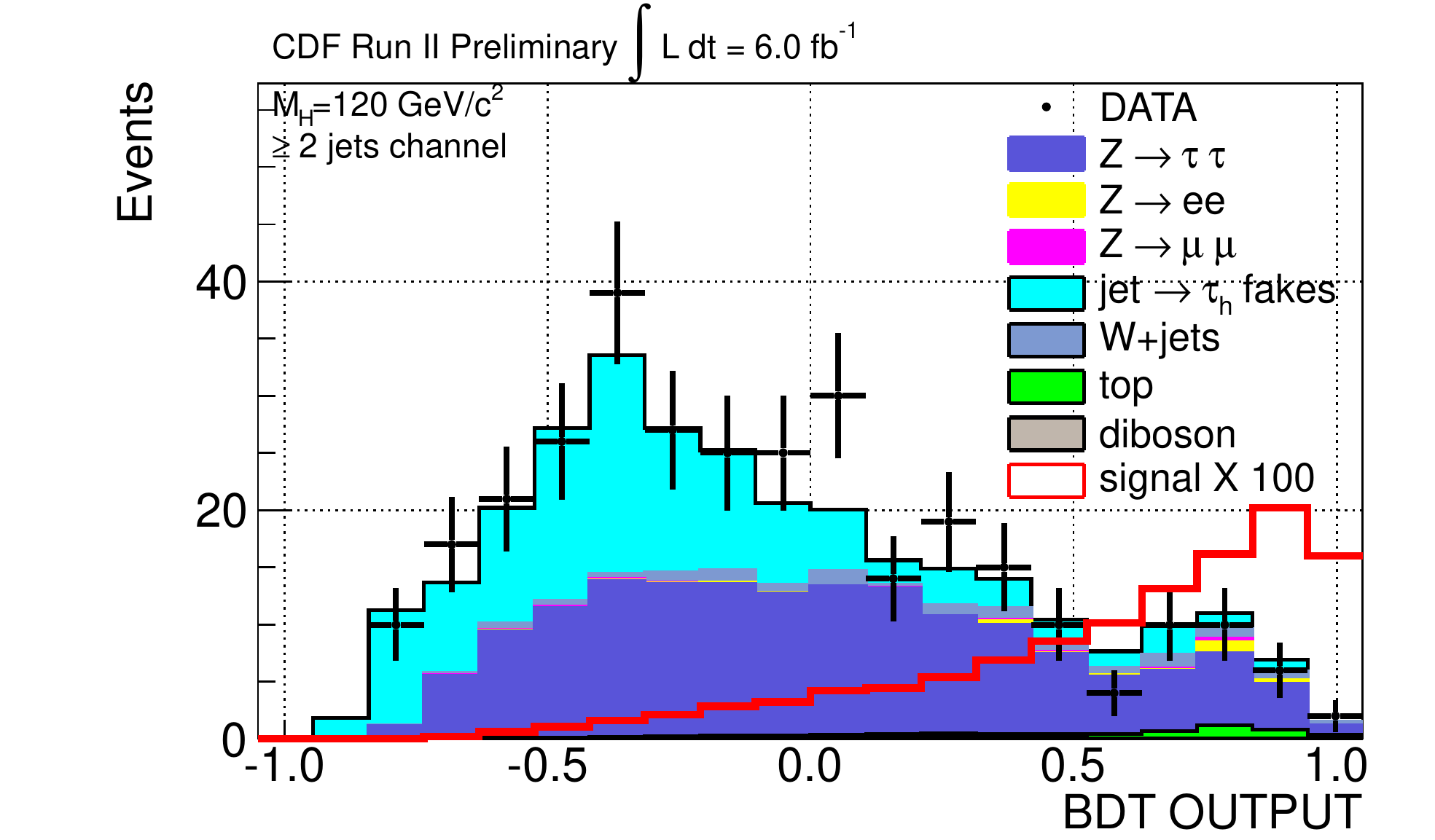}
\includegraphics[width=0.31\textwidth,height=0.15\textheight]{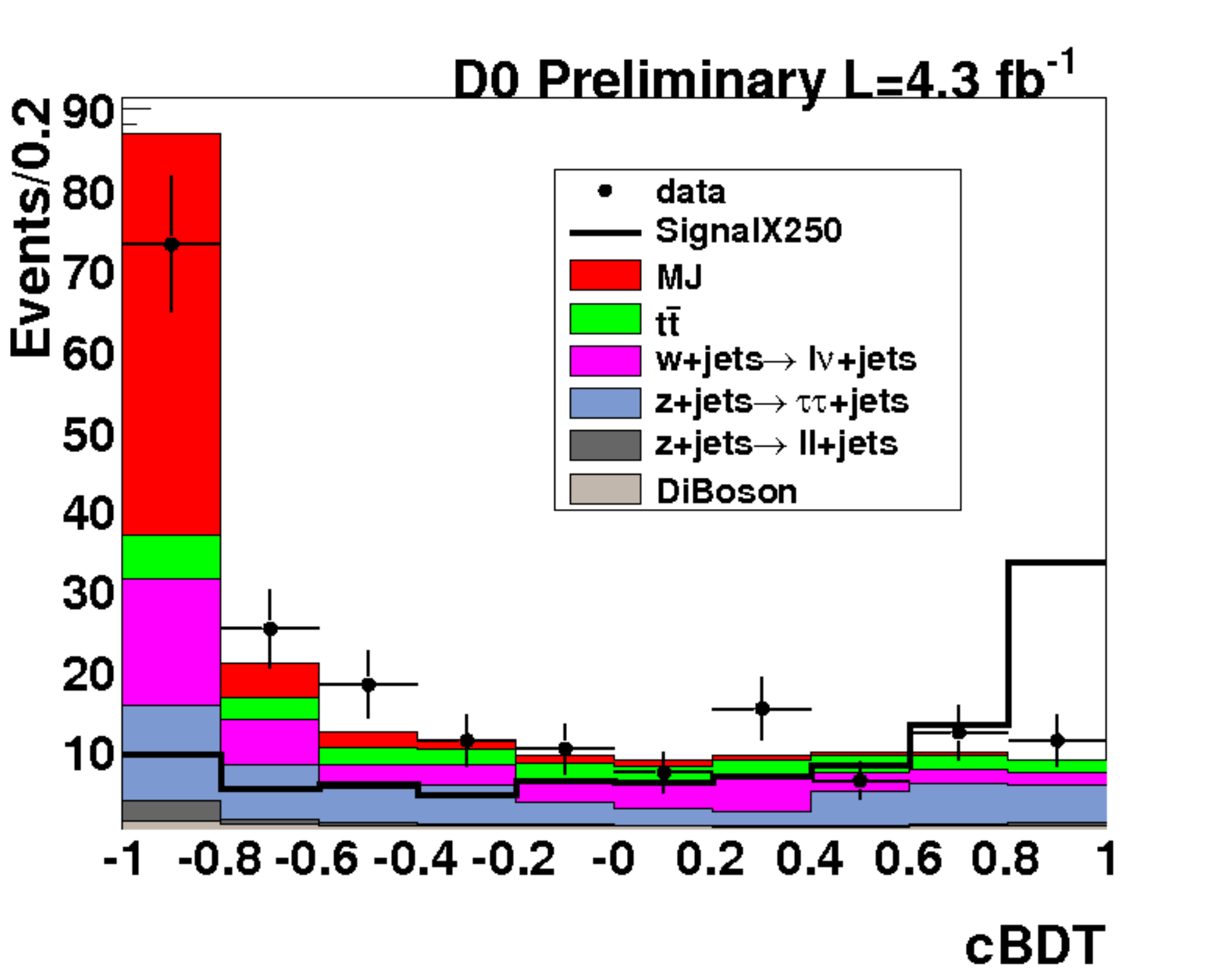}
\makebox[0.31\textwidth][c]{a)}
\makebox[0.31\textwidth][c]{b)}
\makebox[0.31\textwidth][c]{b)}
\caption{a) $\tau-\tau$ invariant mass for the CDF analysis. b) Final discriminant in the 2-jets channel at CDF c) Final discriminant
in the $e-\tau$ channel at D0.}
\label{fig:ditau}
\end{center}
\end{figure}

\begin{figure}[h]
\begin{center}
\includegraphics[width=0.40\textwidth,height=0.18\textheight]{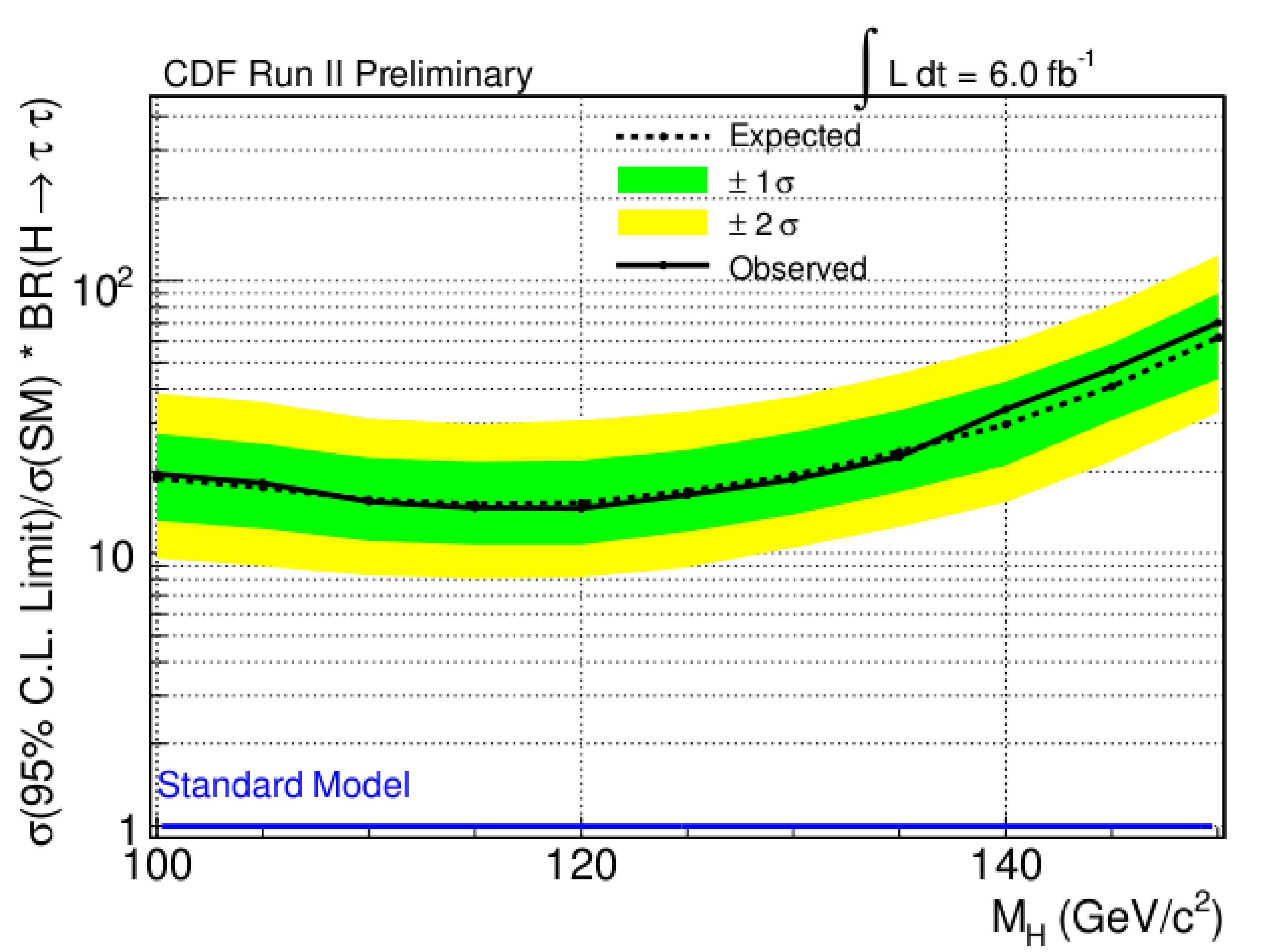}
\includegraphics[width=0.40\textwidth,height=0.18\textheight]{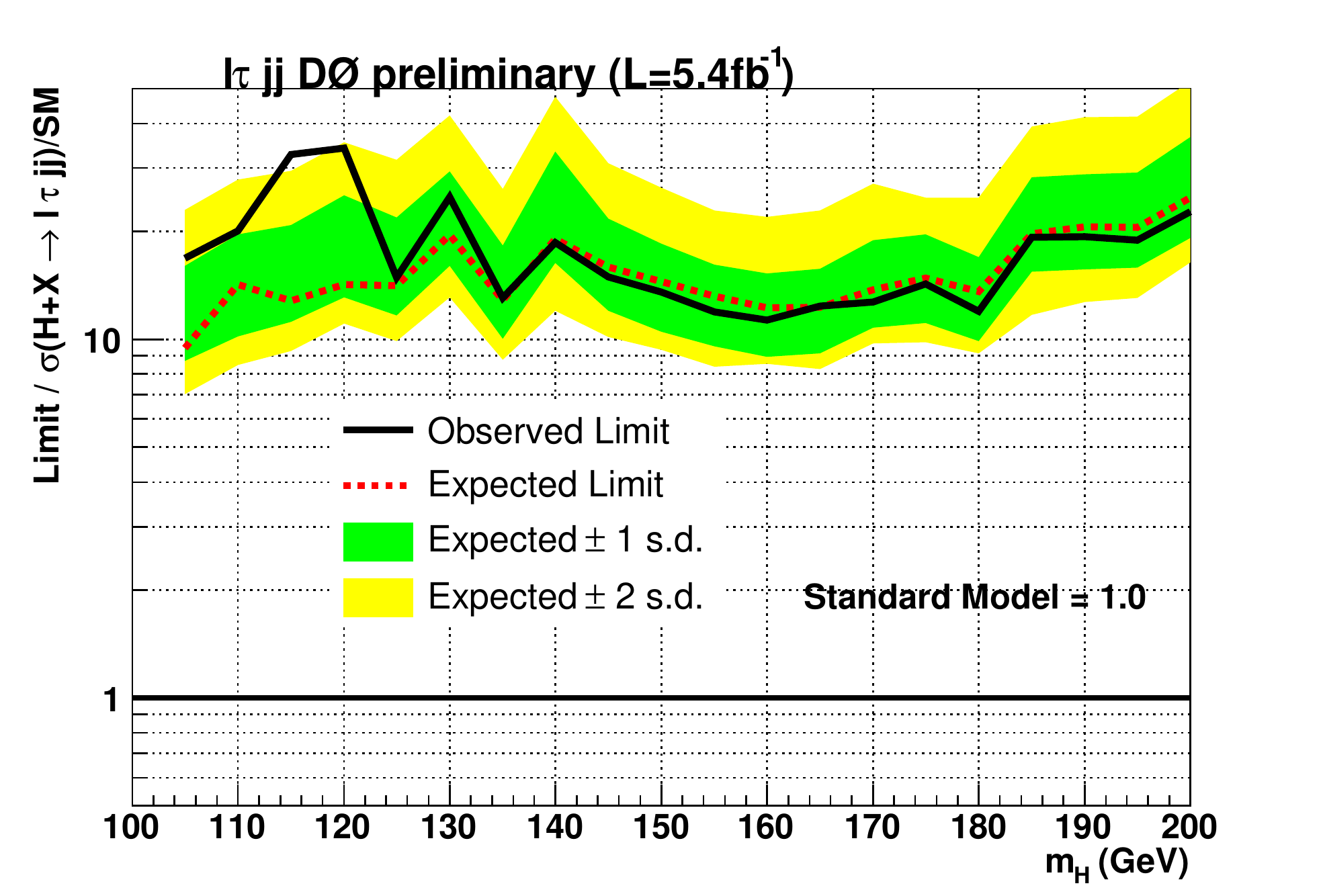}
\makebox[0.40\textwidth][c]{a)}
\makebox[0.40\textwidth][c]{b)}
\caption{95\% C.L. limit on Higgs boson production rate times branching ratio over the Standard Model expection as a function
of the Higgs boson mass for the $H\to \tau \tau$+jets analysis at CDF(a) and D0(b).}
\label{fig:htt_limit}
\end{center}
\end{figure}

Both CDF and D0 search for $H\to\tau_{h}\tau_{l}$, where $l=e,\mu$:
the choice of the ``leptonic+hadronic'' decay final state represents the best compromise
between the high B.R. of the hadronic tau decay and the large background suppression provided by
the request of an electron or a muon. Both the analysis use a multivariate technique to identify the hadronically
decaying $\tau$'s, and distinguish them from jets (see section \ref{sec:tauID} for more details).
Although BR($H \to \tau \tau$) is small, $<10\%$, sensitivity is recovered by including four different
production mechanisms: 
vector boson fusion ($qHq'\to q\tau \tau q'$), that corresponds to 5\% of the total cross section;
associated production with a gauge boson ($WH(\to\tau \tau )$,$ZH(\to \tau \tau)$) that contributes for 18\% of the total cross section; 
gluon fusion ($gg\to H \to \tau \tau$), the main production mechanism, corresponding to 77\% of the total cross section.\\
The $\tau-\tau$ signature alone is not useful to reject the dominant Drell-Yan $Z/\gamma* \to \tau \tau $
background. Due to the presence of at least 3 neutrinos per event and limited energy resolution, 
the $\tau-\tau$ invariant mass has a very broad distribution, and it is not very powerful in separating background from
signal, as Figure \ref{fig:ditau}{\rm a} shows. \\
Both CDF and D0 analyses require the presence of calorimeter jets in the events (one or two jets at CDF, only 2 jets at D0): this 
reduces the size of the signal, and affects the relative contributions of the signal processes.
Events where the Higgs boson is produced through vector boson fusion and in association with a guage boson exhibit kinematical properties
different from the background events. Requiring the presence of jets in the events
increases the contribution of these two processes over gluon fusion, improving the overall sensitivity of the search.
%
A multivariate technique, based on a set of
Boosted Decision Trees, is used to maximally exploit the kinematical differences between the $H \to \tau \tau$ signal
and the backgrounds. We do not see evidence for a Higgs boson signal and therefore we set a 95\% confidence limit
(C.L.) upper limit on the Higgs boson production cross section times BR($H\to \tau\tau$) relative to the Standard Model (S.M.) prediction.
Figure \ref{fig:htt_limit}{\rm b/c} show the results for a Higgs boson mass varying from $m_{H} = 100 \ \gevcc$ to $m_{H} = 150\ \gevcc$ 
for the CDF analysis and up to  $m_{H} = 200\ \gevcc$ for D0. The analysis performed at D0 is sensitive to a
wider range of masses, since it also considers signal events from $H \to W^{+} W^{-}$. \\
CDF observes a 95\% C.L. limit of 14.7 times S.M. prediction ($m_{H} =115 \ \gevcc$), while expecting 
$15.2^{+6.5}_{-4.4}$. For the same test mass, D0 analysis observes a limit of 32.8 times S.M., while expecting 12.8.
More details can be found in \cite{htt_cdf} and \cite{htt_d0}.

\subsection{Higgs Associated production with W/Z}
\begin{figure}[h]
\begin{center}
\includegraphics[width=0.31\textwidth,height=0.15\textheight]{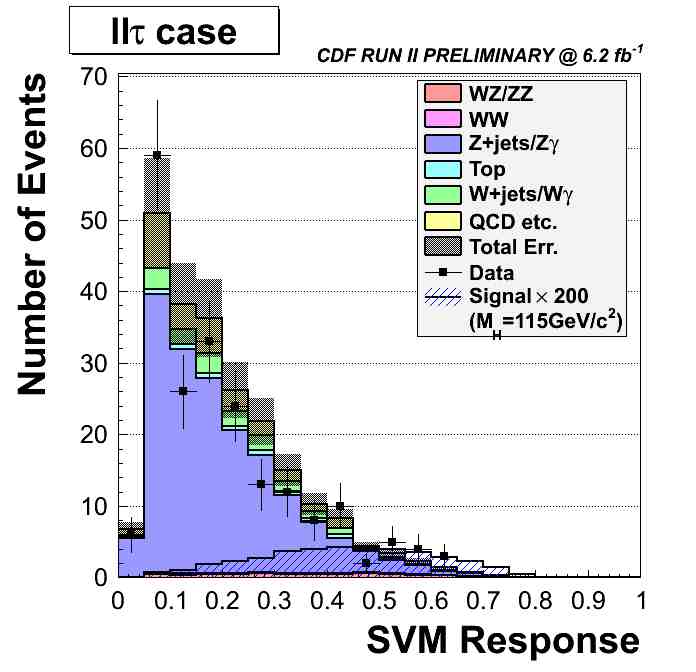}
\includegraphics[width=0.31\textwidth,height=0.15\textheight]{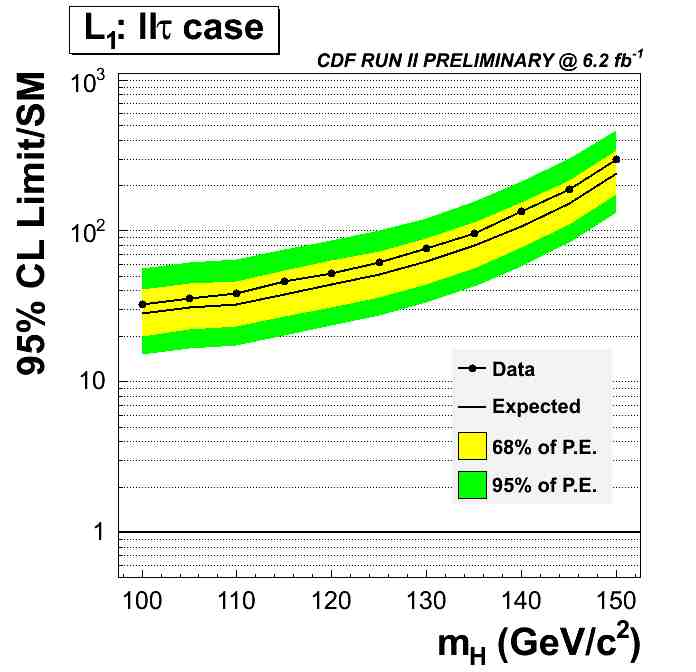}
\includegraphics[width=0.31\textwidth,height=0.15\textheight]{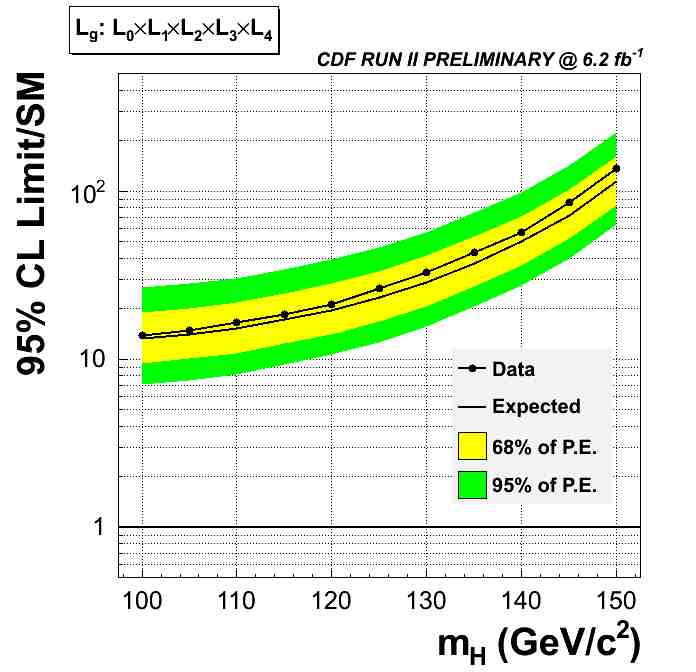}
\makebox[0.31\textwidth][c]{a)}
\makebox[0.31\textwidth][c]{b)}
\makebox[0.31\textwidth][c]{b)}
\caption{a) Comparison between the background model of the final discriminant and the data distribution, in the $ll\tau_{h}$ channel. 
b)95\% C.L. upper limit on the Higgs boson production rate divided by the standard model expectation as a function of the
Higgs boson mass in the $ll\tau_{h}$ channel.  
c) 95\% C.L. upper limit on the Higgs boson production rate divided by the standard model expectation as a function of the Higgs boson mass,
for all the channels combined. 
}
\label{fig:lltt_SVM}
\end{center}
\end{figure}

We present here a search for a Higgs boson produced in association with a $W/Z$ boson,
in final states $WH\to l \nu \tau \tau$ and $ZH\to ll \tau \tau$, where $l = e,\mu$. This analysis, presented for the first
time at the 2011 summer conferences, considers all the possible decay modes of the $\tau$. The search analyzes 5 different
independent channels: $lll$, $ll\tau_{h}$, $e\mu\tau_{h}$, $l\tau_{h}\tau_{h}$, $LLLL$, where $l=e,\mu$ and $L=e,\mu,\tau_{h}$. The leptons
are identified using standard rectangular cuts (see \cite{lltt} for more details).
Only a minimal set of requirements are applied in the selection of the events, beside the presence of the leptons.
One requirement is that the sum of the lepton charges is consistent with the presence of a $W$ or a $Z$ boson 
( $\vert \sum Q \vert = 1$ for tri-lepton events, and $\vert \sum Q \vert = 0$ for four-leptons events).
Multiple neutrinos from the $\tau$ decay are present in the signal events:
requiring that the missing transverse energy significance satisfies $\met^{Sig} = \met / \sqrt{\sum \et} \ge 1$ reduces
the contributions of backgrounds where no real $\met$ is present, like Drell-Yen events and QCD mulitjet production. \\
A Supplement Vector Machines (SVM) is used to separate signal from backgrounds in candidate events. A SVM \cite{SVM} is a supervised
training binary classifier, effective when trained on low statistic samples. The idea of an SVM is to determine the hyperplane
in the space of the training variables that separates maximally two classes of events.
Depending on the channel, different SVMs are trained to separate the signal from each of the backgrounds, and their individual output 
scores are multiplied together: this defines the final discriminant in Figure \ref{fig:lltt_SVM}{\rm a}.\\
All the tri-leptons channels containing at least one $\tau_{h}$ have very similar sensitivity: the $ll\tau_{h}$ channel, for example,
is sensitive to a Higgs boson ($m_{H}= 115 \ \gevcc$) production rate of 38.2 times the S.M. expected one, see \ref{fig:lltt_SVM}{\rm b}.
The best sensitivity is achieved combining all the channels, as shown in Figure \ref{fig:lltt_SVM}{\rm c}:
at $m_{H} = 115 \ \gevcc$, the observed limit divided by the the Standard Model prediction is 18.5, 
while the expected one is $17.3^{+7.4}_{-3.9}$.
\cite{htt_d0}.

\section{Including $\tau$ leptons into main searches}
With Tevatron data-taking coming to an end, both CDF and D0 collaborations, in a effort to maximize their
sensitivity to the Higgs boson, are pursuing more challenging channels. The main analyses, $WH\to l \nu b\bar{b}$ at
low mass, and $H\to WW \to l \nu l \nu$ at high mass, target events with $e$ and $\mu$.
Adding final states with hadronically decaying $\tau$'s is a natural step.
The higher amount of backgrounds typical of $\tau$ final states, though, limits the sensitivity of these analyses.
%

\subsection{$H \to WW$ with $\tau$ leptons in final state}
\begin{figure}[h]]
\begin{center}
\includegraphics[width=0.31\textwidth,height=0.15\textheight]{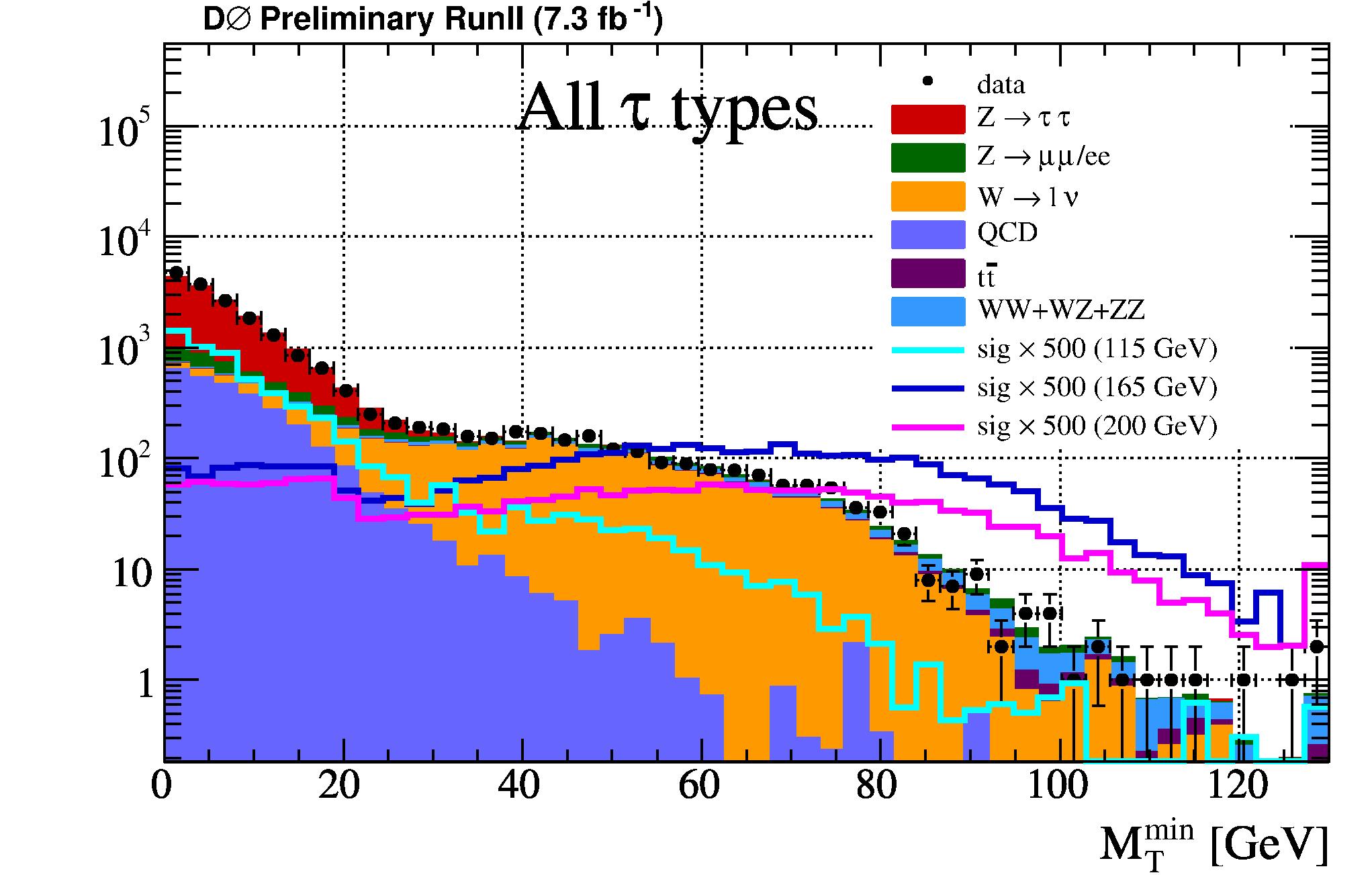}
\includegraphics[width=0.31\textwidth,height=0.15\textheight]{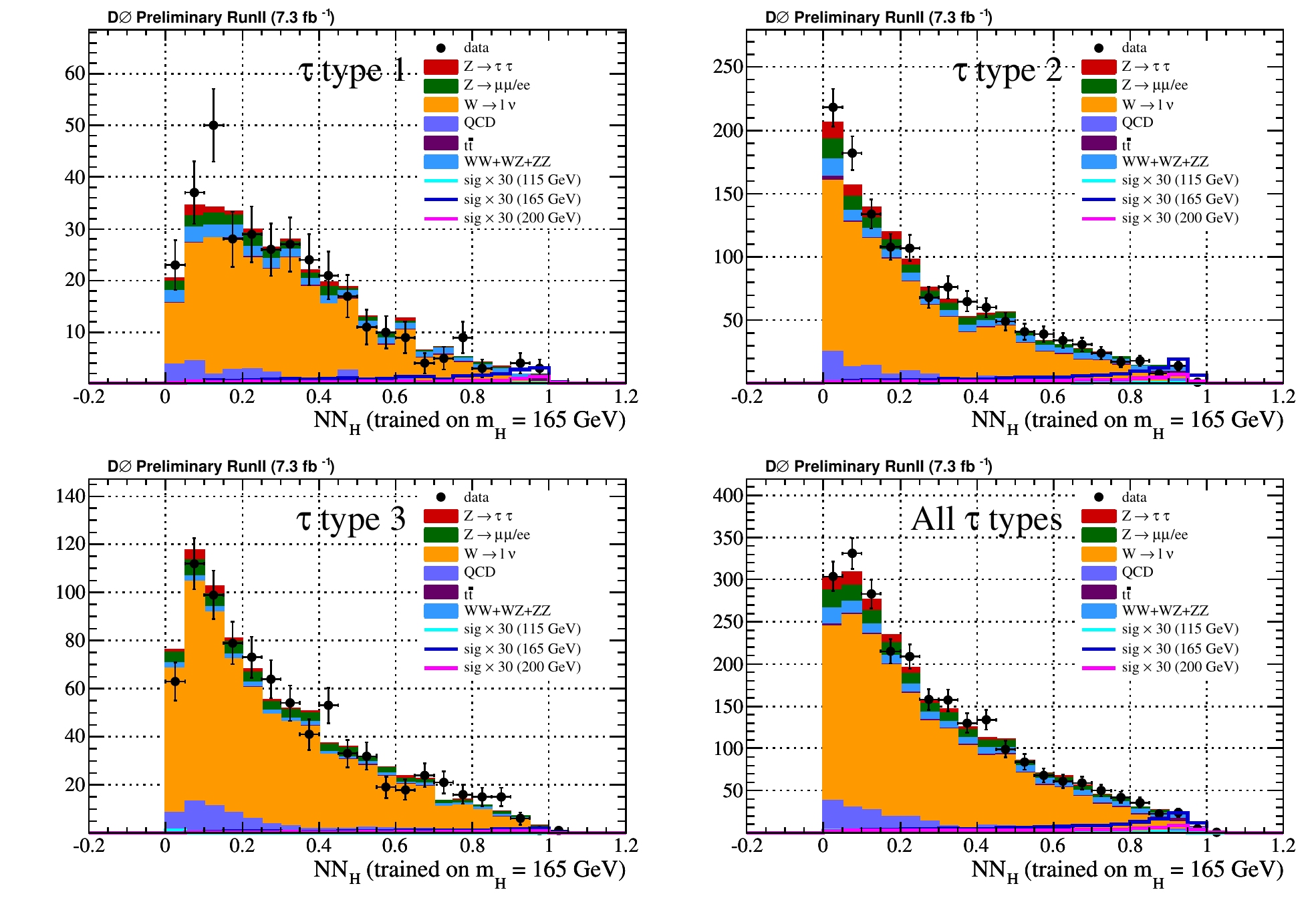}
\includegraphics[width=0.31\textwidth,height=0.15\textheight]{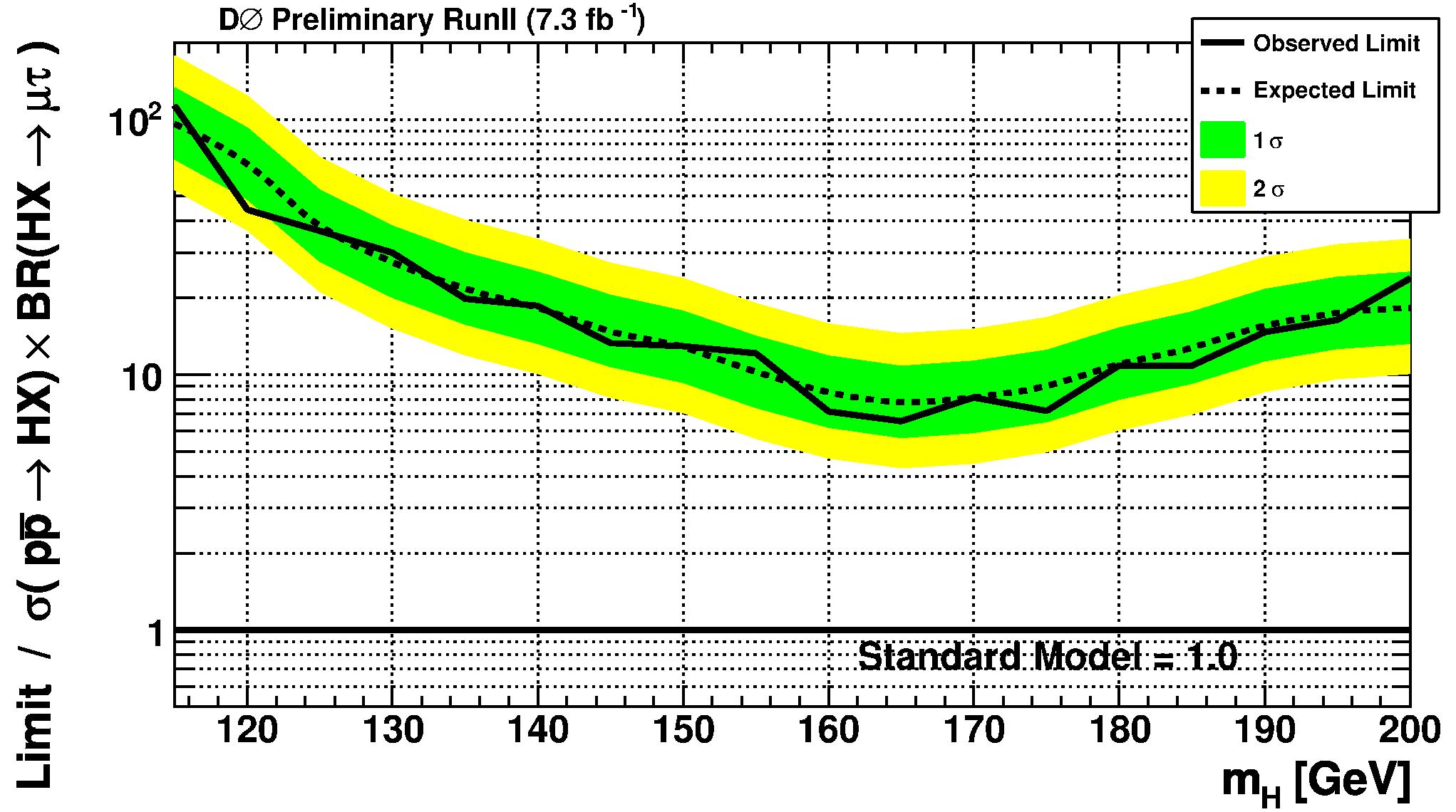}
\makebox[0.31\textwidth][c]{a)}
\makebox[0.31\textwidth][c]{b)}
\makebox[0.31\textwidth][c]{c)}
\caption{D0 analysis. a) Transverse mass distribution in $\mu-\tau$ events. The Drell-Yen and multijet backgrounds, where
$\met$ arise from the energy mis-measurement of one of the object in the event, pile up at lower values of $m_{\rm T}$.
Boosted Decision Tree output distribution for events in signal sample, in the $\mu-\tau$ sample.
b) Boosted Decision Tree output distribution for events in signal sample.
c)95\% C.L. upper limit on the Higgs boson production rate time BR($H \to WW$) as a function of the Higgs boson mass.}
\label{fig:hww_d0}
\end{center}
\end{figure}

\begin{figure}[h]
\begin{center}
\includegraphics[width=0.31\textwidth,height=0.15\textheight]{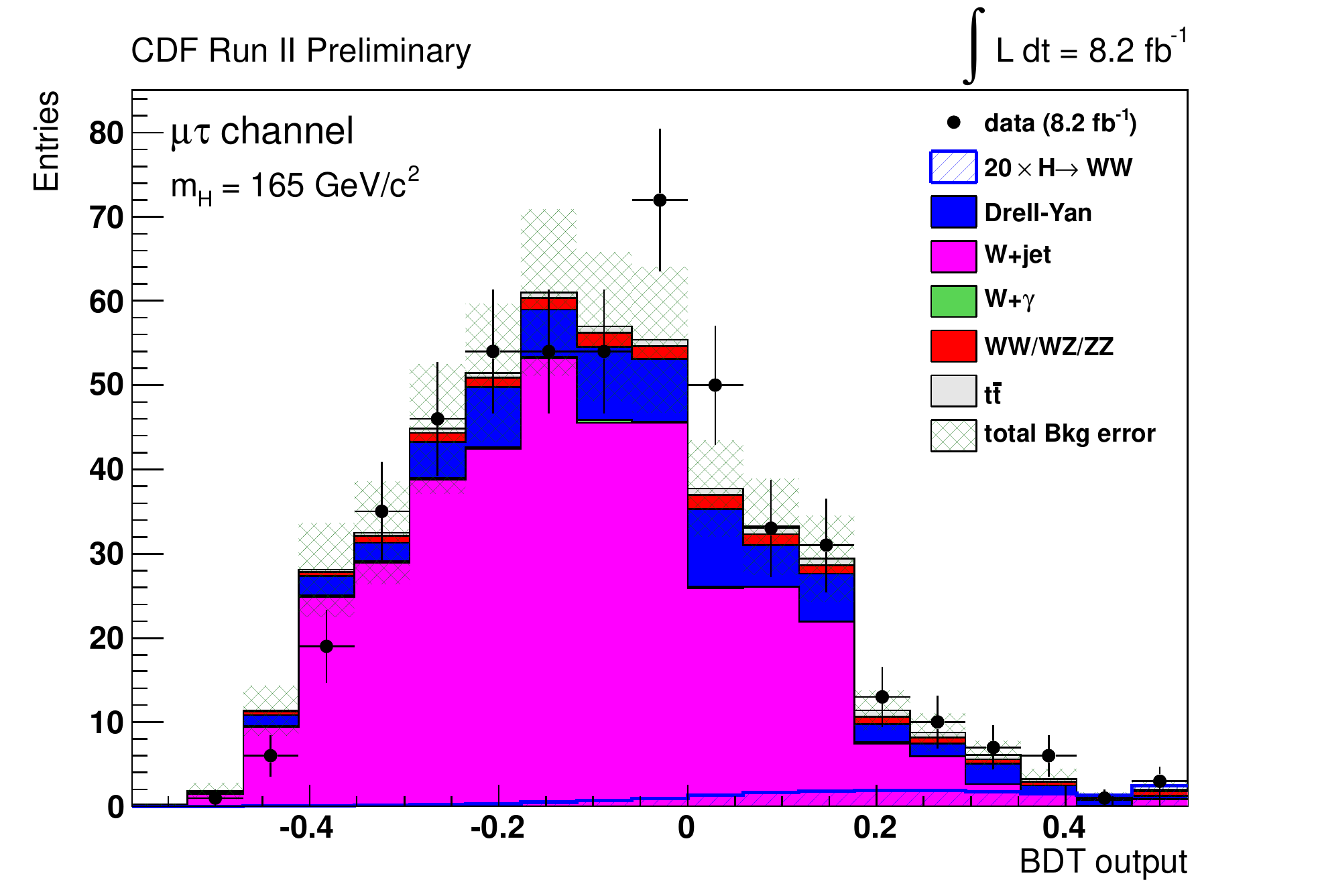}
\includegraphics[width=0.31\textwidth,height=0.15\textheight]{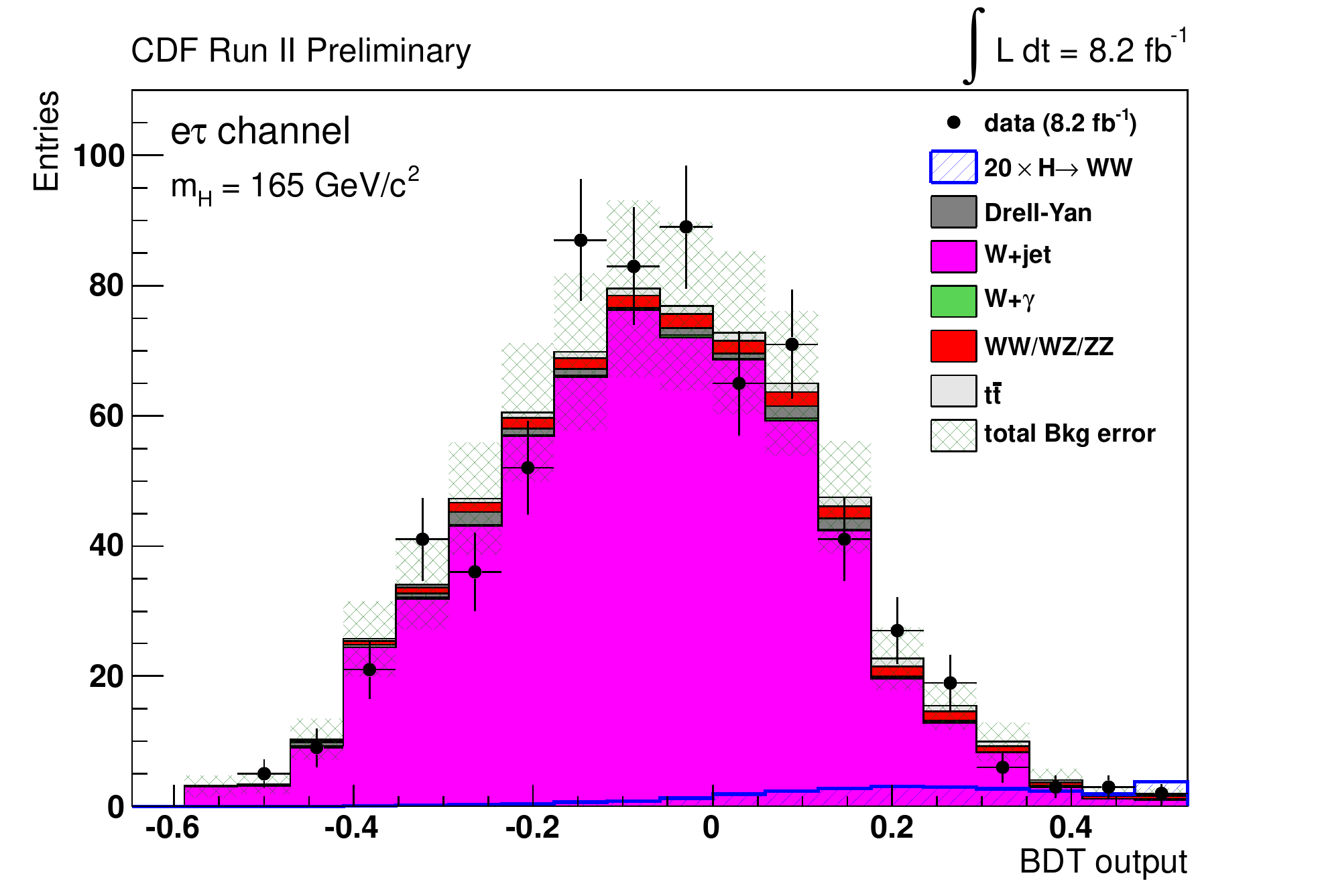}
\includegraphics[width=0.31\textwidth,height=0.15\textheight]{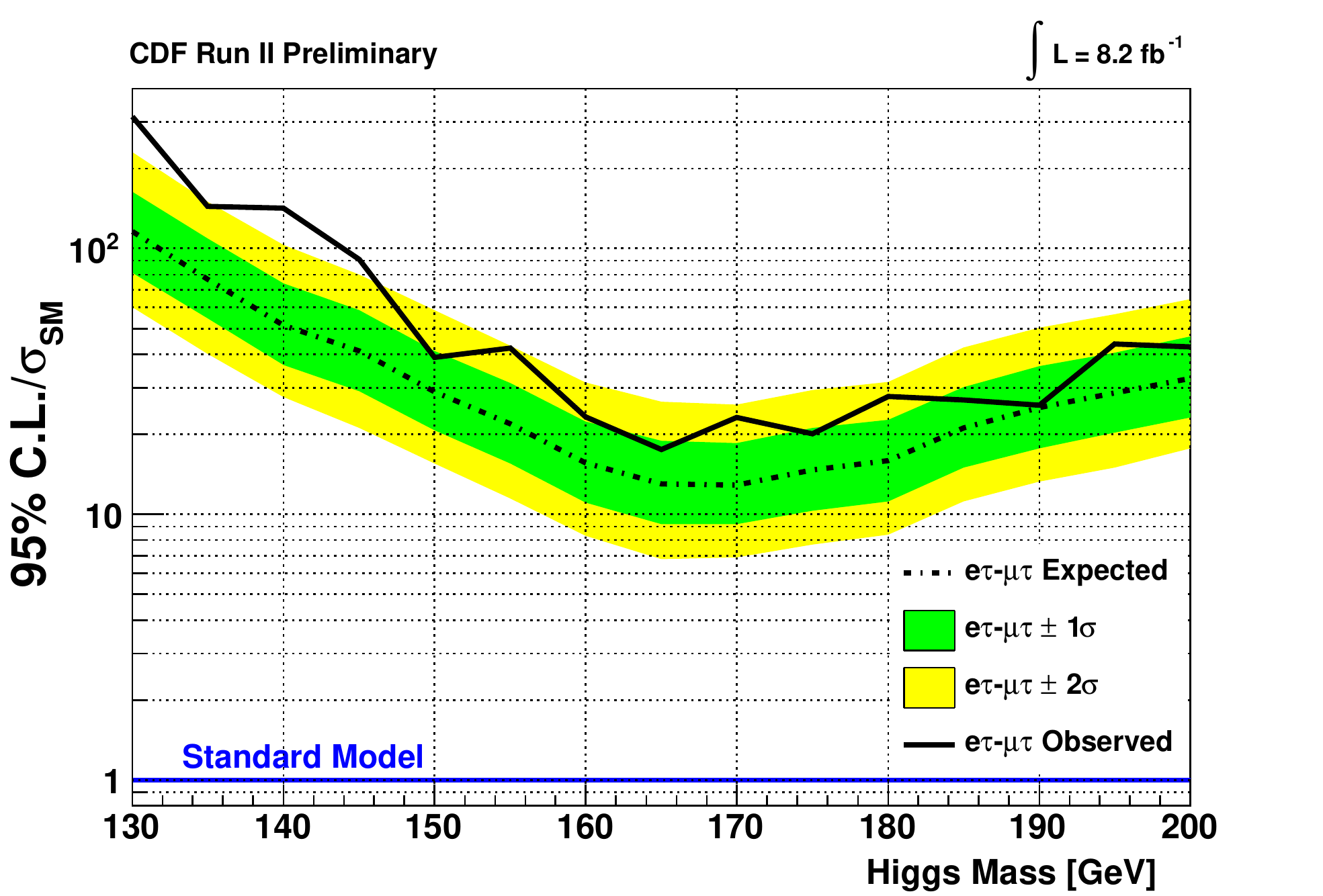}
\makebox[0.31\textwidth][c]{a)}
\makebox[0.31\textwidth][c]{b)}
\makebox[0.31\textwidth][c]{b)}
\caption{CDF analysis a) Boosted Decision Tree output distribution for events in signal sample, in the $\mu-\tau$ sample.
b) Boosted Decision Tree output distribution for events in signal sample, in the $e-\tau$ sample.
c)95\% C.L. upper limit on the Higgs boson production rate time BR($H \to WW$) as a function of the Higgs boson mass for the two channels combined.}
\label{fig:hww_cdf}
\end{center}
\end{figure}

We present here the search for $gg \to H \to WW \to l \nu \tau\nu $, where the $\tau$ decays hadronically.
Final states where the $\tau$ decays leptonically are already included in the main $H \to WW$ analysis with $e$ and $\mu$.
The CDF collaboration considers events triggered on the presence of either a central $e$ or a $\mu$: this sets the lower bound on the
lepton $\pt$ ( $\pt \ge \ 20 \GeV$). The $\tau$ signal cone of aperture $\alpha = min(0.17, 5/E^{clu} [\GeV]$) rad 
is built around a track with $\pt \ge 10 \ \gevc$ (seed track), where $E^{clu}$ is the calorimeter energy of the $\tau$-candidate.
To be identified, the $\tau$ must exhibit $E_{T}^{Vis} \ge 15 \ (20) \ \GeV$ for 1-prong(3-prong) candidates, be 
central ($\vert \eta\vert \le 1$, and be isolated.\\
The D0 collaboration, instead, selects only events in the ``muon+tau'' final state, using a mixture of single and dilepton
triggers requiring both lepton and jet signatures. The $\mu$'s are identified down to a $\pt \ge 15 \ \gevc$, and
must have pseudorapidity $\vert \eta \vert \le 1.6$. The hadronically decaying $\tau$, built around a 
a seed track with $\pt \ge 7/5/7 \ \gevc$, must have have a transverse momentum,
as measured in the calorimeter, $\pt \ge 12.5/12.5/15\ \gevc$ for type-I/II/III, and $\vert \eta \vert \le 2.0$.
The $\tau$-candidate, not to be rejected, mush exhibit a neural network score $NN_{\tau}\ge0.9/0.9/0.95$.
This selection corresponds to an overall efficiency of $\sim$ 55\% in signal events for a fake rate of $\sim$ 2\% in multijet events.\\
\indent
The sample where the two leptons are required to have opposite charge is dominated by Drell-Yen $Z/\gamma* \to \tau\tau$ events, with 
contribution from W+jets and multijet production. 
The Drell-Yen and multijet backgrounds can be easily suppressed: 
D0 requires transverse mass $m_{T}(l,\met) = \sqrt{2\vert \pt^{l} \vert \vert \met \vert (1 -cos\Delta\Phi(l,\met)} \ge 25 \gevcc$ (see
Figure \ref{fig:hww_d0}{\rm a}), while
CDF cuts on different kinematical variables, such as the $\met$, the angle between the dilepton system and the $\met$. 
More details are described in \cite{HWW_cdf}.\\
The dominant background after these selections is the W production when the gauge boson decays into $e$ or $\mu$, and an additional jet 
is identified as a $\tau$ candidate. Monte Carlo simulated events are used to describe it. Both the collaboration developed corrections to
the $\tau$ mis-identification rate, that is not properly modeled in simulated events.\\
There is no single variable able to distinguish the $W$+jets background from the signal, so both the collaborations use multivariate techniques
to maximize the separation between the two processes.
One of the most powerful variable is the angular correlation of the leptons, due to the spin of the Higgs particle.\\
CDF adds $\tau$-related observables, as the energy deposited in the isolation annulus, to the kinematical 
variables used as inputs to the Boosted Decision Tree. More details can be found in \cite{HWW_cdf}.\\
Figure \ref{fig:hww_d0}{\rm b}  shows the final Neural Network discriminant in the D0 analysis for the different classes of $\tau$'s separately, 
and for all of them combined. Details on the input variables to the NN discriminant can be found in \cite{HWW_d0}.\\
Figure \ref{fig:hww_d0}{\rm c} shows the 95\% C.L. limit of the Higgs boson production rate time BR($H \to WW$)
as a function of its mass. For $m_{H} = 160 \ \gevcc$, the observed(expected) limit is 7.2(6.5) time the S.M. expectation.\\ 
Figures \ref{fig:hww_cdf}{\rm a} and \ref{fig:hww_cdf}{\rm b} show, for the CDF analysis, the final 
Boosted Decision output for the $e-\tau$ and $\mu-\tau$ final states, and the combined 95\% C.L. limit, fig \ref{fig:hww_cdf}{\rm c}.
For $m_{H}=160 \ \gevcc$, the observed(expected) 
limit is 23.3(15.6). Among the two CDF channels, the $e-\tau$ one has the best sensitivity (with an expected limit of 20.5 time the S.M. expectation 
for $m_{H}=160 \ \gevcc$): the $\mu-\tau$ events have a sensitivity $\sim$ 20\% worst than that.

\subsection{WH associated production with $\tau$'s}
\begin{figure}[h]
\begin{center}
\includegraphics[width=0.31\textwidth,height=0.15\textheight]{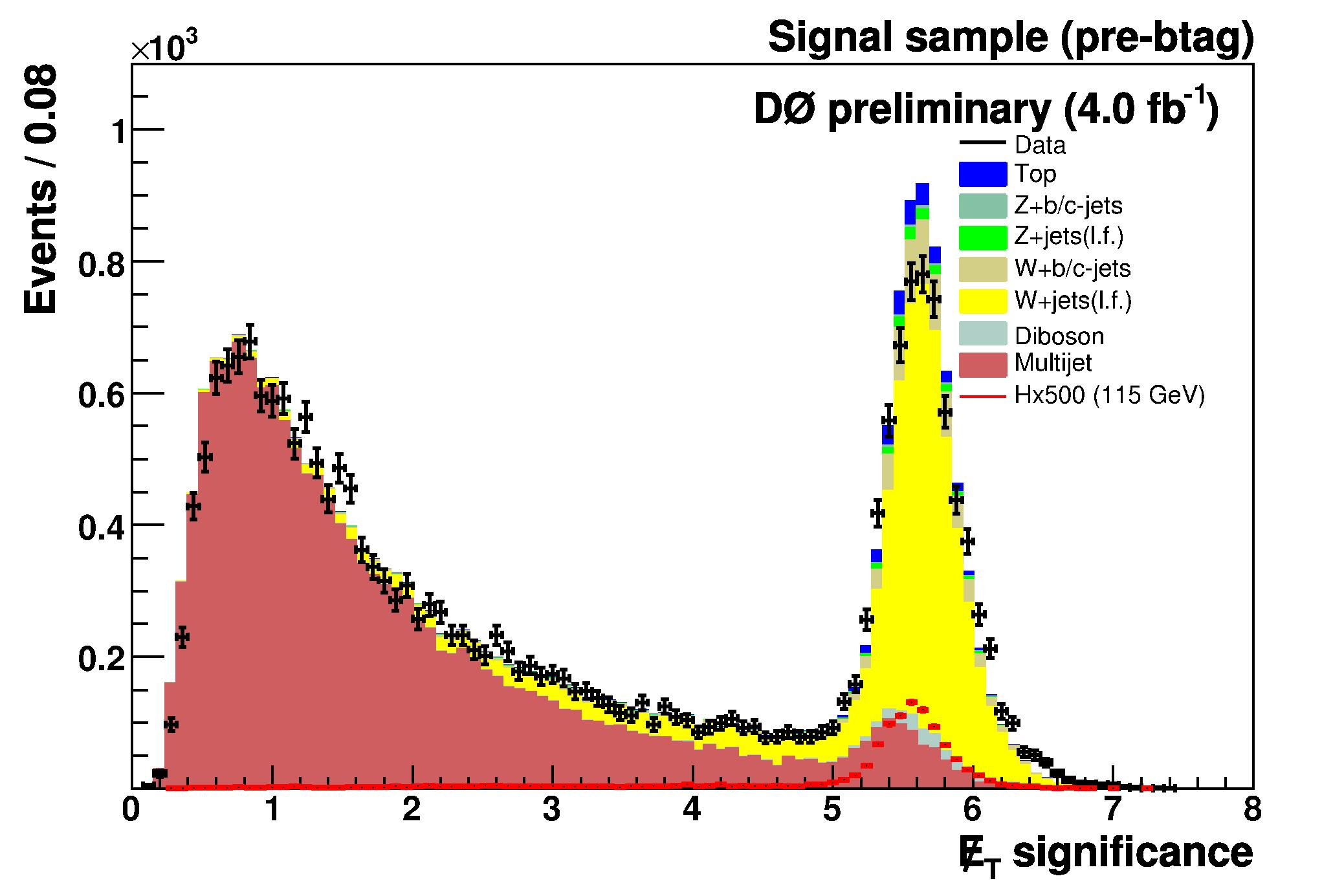}
\includegraphics[width=0.31\textwidth,height=0.15\textheight]{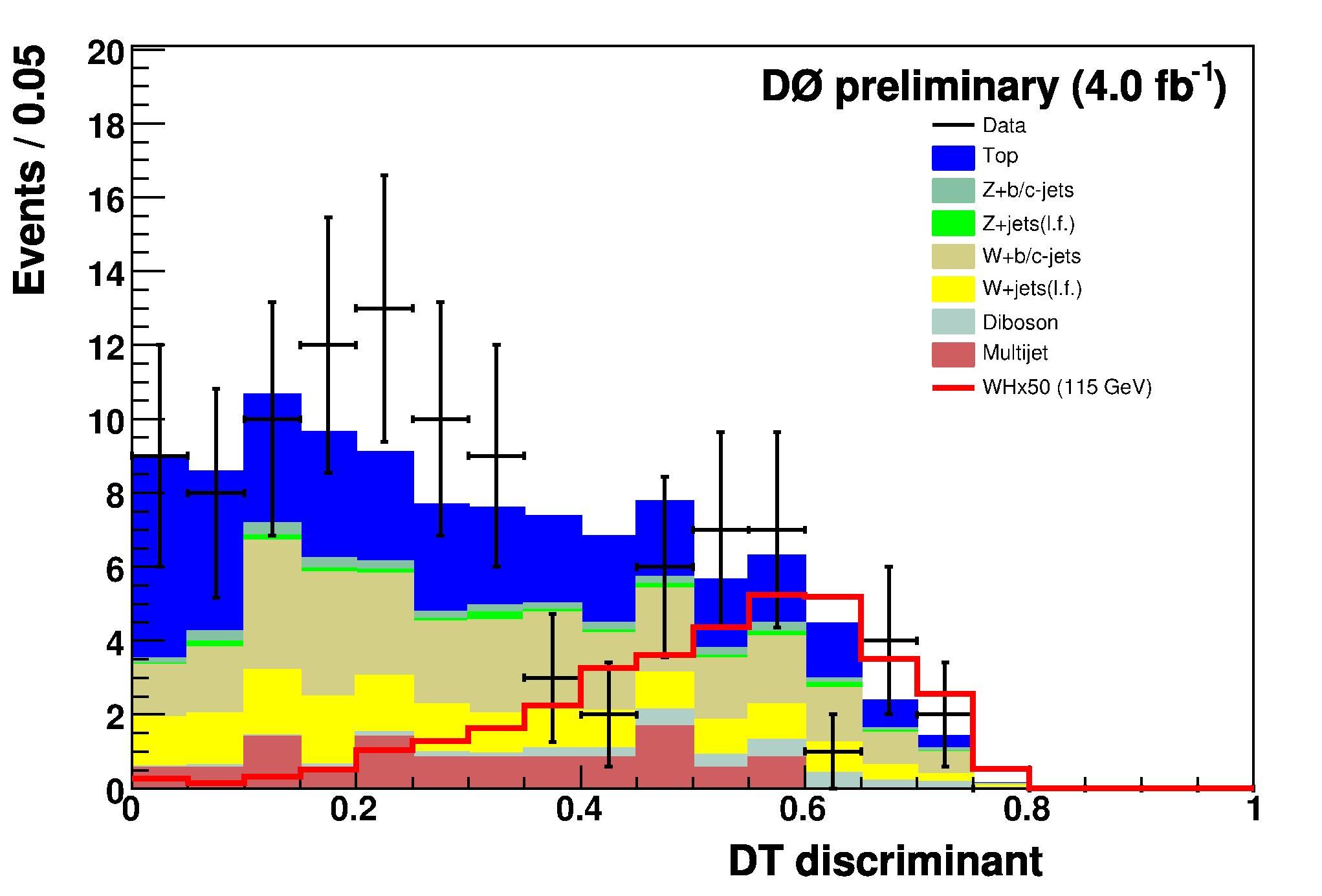}
\includegraphics[width=0.31\textwidth,height=0.15\textheight]{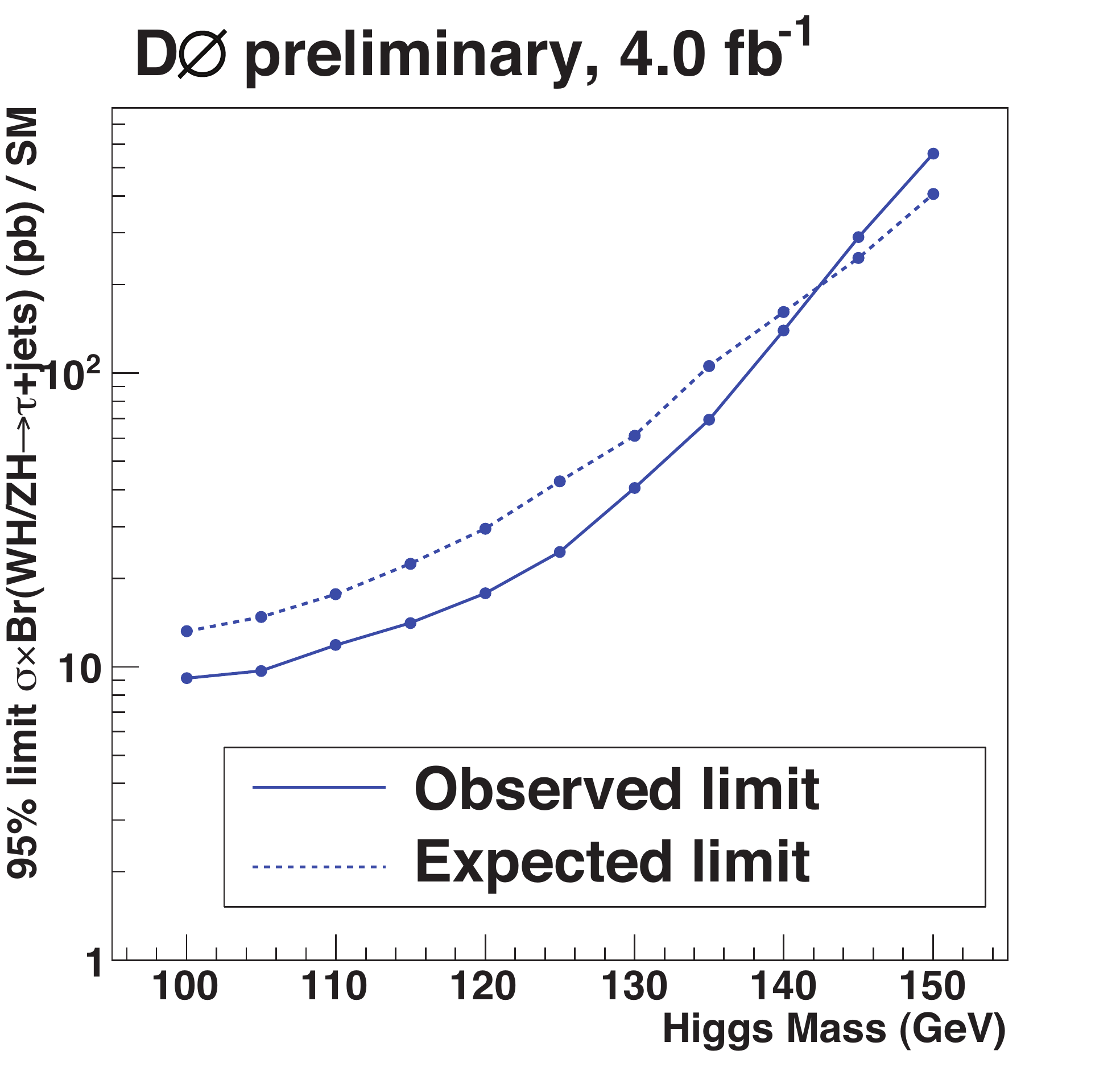}
\makebox[0.31\textwidth][c]{a)}
\makebox[0.31\textwidth][c]{b)}
\makebox[0.31\textwidth][c]{b)}
\caption{a) D0 analysis: $\met^{sign}$ distribution for events with an identified type-II $\tau$ and exactly two jets, before requiring $b$-tagging.
b) BDT output for events with one type-II $\tau$ and two jets, both $b$-tagged. The red histograms represent the distribution for signal events, 
multiplied by a factor of 50.
 c)Final 95\% C.L. limit on the Higgs boson productions in factors away from the S.M. prediction.}
\label{fig:whtnbb_d0}
\end{center}
\end{figure}

\begin{figure}[h]
\begin{center}
\includegraphics[width=0.31\textwidth,height=0.15\textheight]{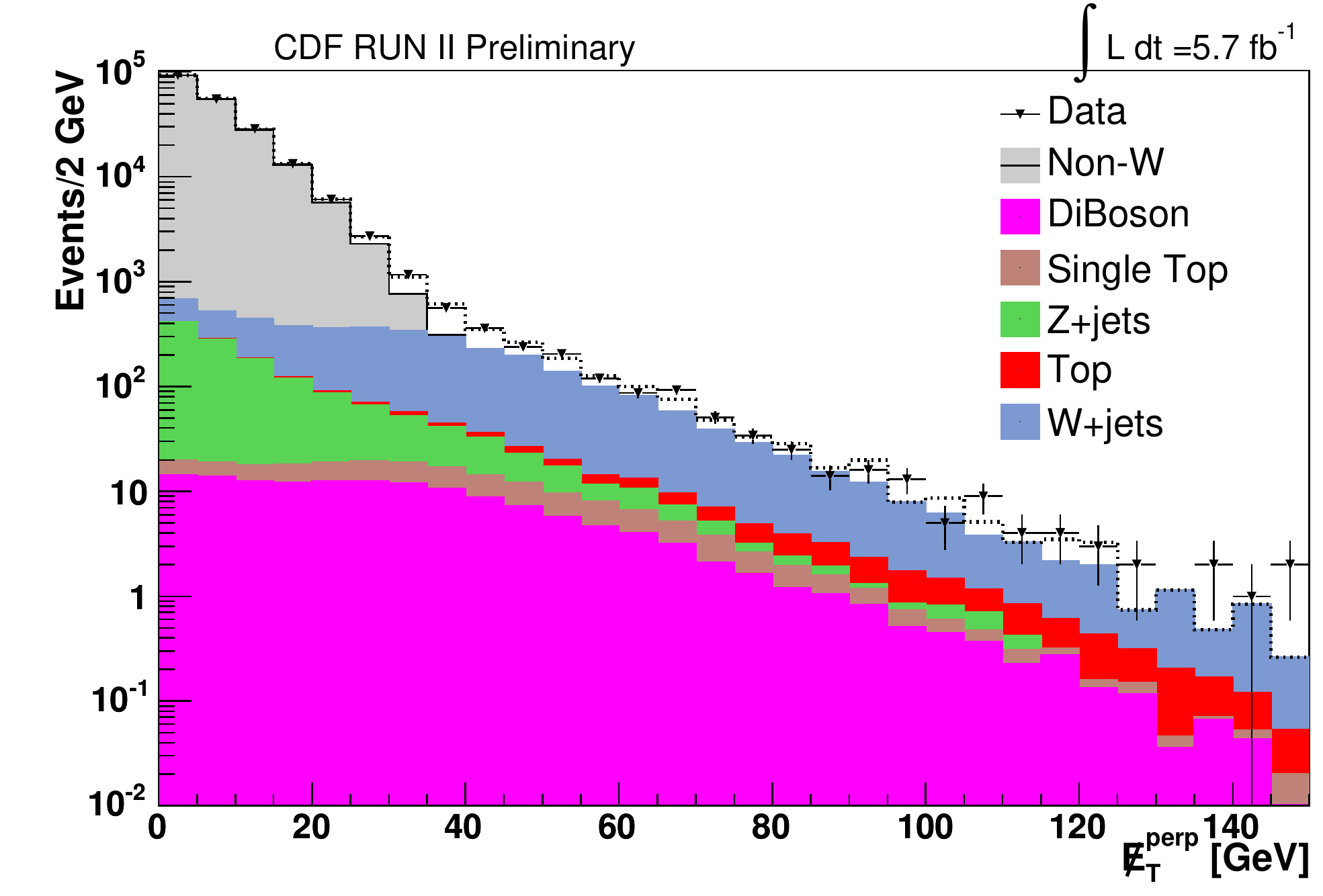}
\includegraphics[width=0.31\textwidth,height=0.15\textheight]{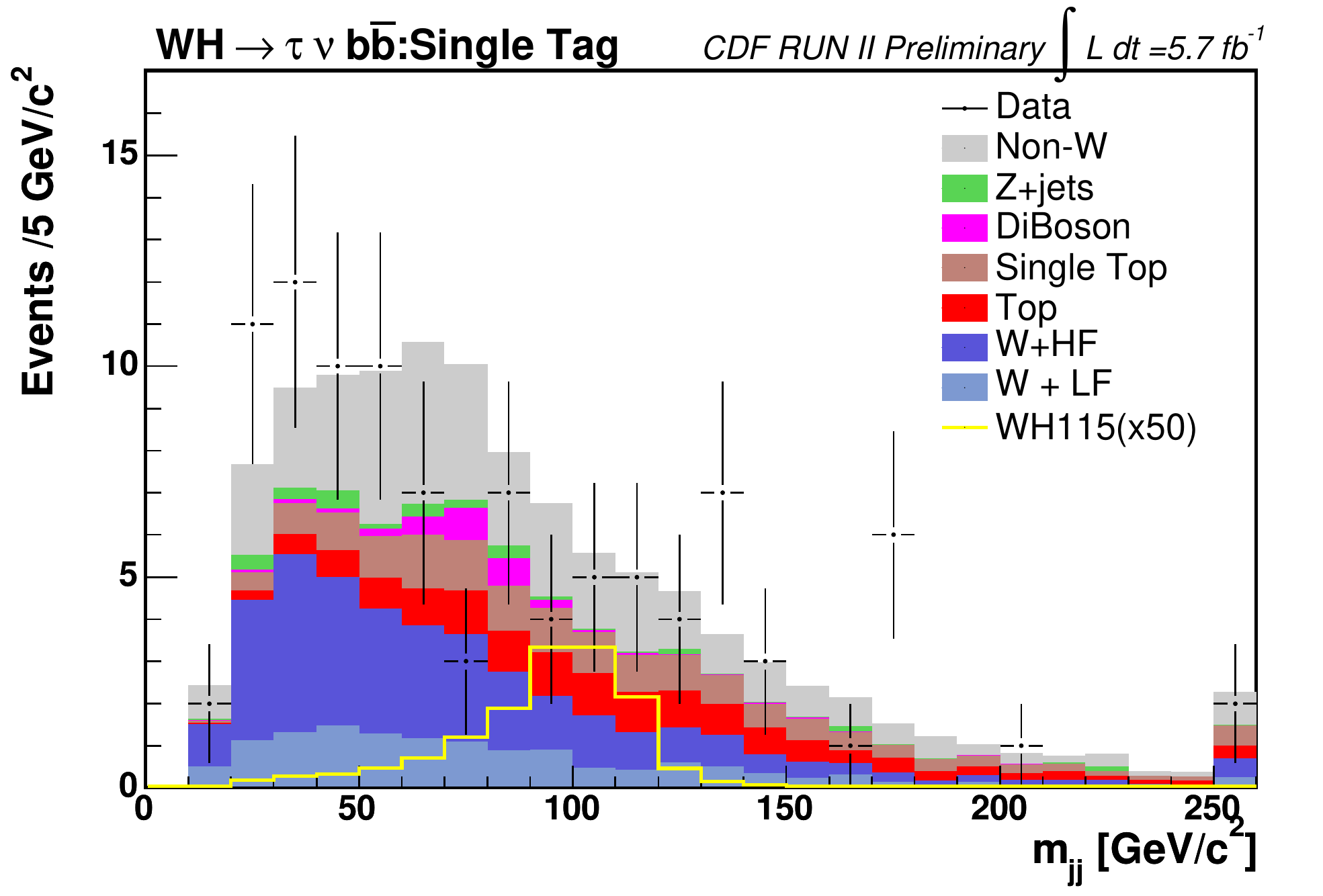}
\includegraphics[width=0.31\textwidth,height=0.15\textheight]{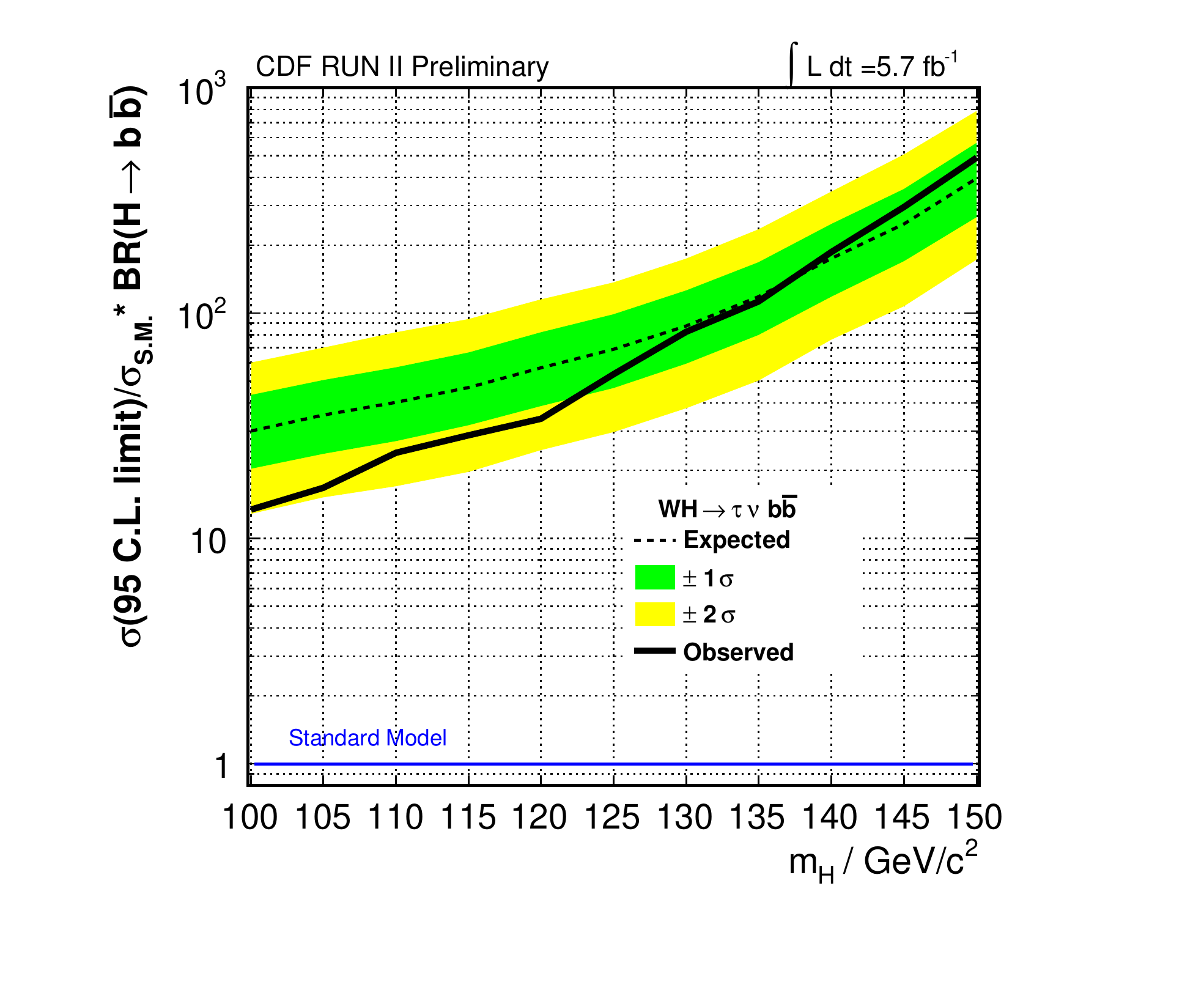}
\makebox[0.31\textwidth][c]{a)}
\makebox[0.31\textwidth][c]{b)}
\makebox[0.31\textwidth][c]{b)}
\caption{a) CDF analysis: $\met^{perp}$ distribution for events with an identified $\tau$ and exactly two jets, before requiring $b$-tagging.
The dash histograms represent the sum of the background contributions.
b)Di-jet invariant distribution for events in signal sample with exactly one $b$-tagged jets. The yellow line represents the Higgs boson signal ($m_{H} =115 \ \gevcc$)
time 50. c) 95\% C.L. limit on the Higgs boson productions in factors away from the S.M. prediction. The limits set in events with exactly one $b$-tagged jet
and two $b$-tagged jets are combined to give the final result.}
\label{fig:whtnbb_cdf}
\end{center}
\end{figure}

We detail here a search for a low mass Higgs boson, when produced in association with a W boson, in the $WH \rightarrow \tau \nu b \bar{b}$ final state.
The final state topology considered in this analysis is a pair of b-jets from the decay of the Higgs boson, with missing
transverse energy $\met$ and a hadronically decaying $\tau$ lepton ($\tau_{h}$ ).\\
The events analyzed are recorded using a trigger that selects events with jets and $\met$ at D0, and one that selects events with a $\tau$ candidate and
$\met$ at CDF.
The choice of trigger affects the event selection: in the D0 analysis, the type-I and type-II $\tau$ candidates must exhibit $\pt^{seed} \ge 7/5  \ \gevc$, 
and transverse momentum, measured using calorimeter information only, $\pt^{Clu} \ge 12/10 \ \gevc$. A Neural Network algorithm is
used to reduce the contamination of jets mimicking the $\tau$ signature ($NN_{\tau}\ge 0.9$), retaining 65\% of the hadronically decaying $\tau$'s.
Since the trigger used in the CDF analysis requires the presence of a $\tau$ candidate, the offline cuts used to identify the $\tau$ must be at least as tight
as the ones used online. The $\tau$ candidate, reconstructed with a fixed-size cone with aperture $\alpha = 10\,^{\circ}$, must satisfy $\pt^{seed} \ge 10 \ \gevc$
and $\pt^{Clu} \ge 20 \ \gevc$, and must be isolated. The efficiency for this selection is $\sim$ 35\%: most of the inefficiency is due to the 
tight energy requirements. More details can be found in \cite{whtnbb_d0} and \cite{whtnbb_cdf}.\\
The main backgrounds arise from (W/Z)bb, from (W/Z )+(non-b jets) due to flavor mis-identification (mis-tagging), from top quark production, diboson production,
and from multijet events with fake $\met$ resulting from fluctuations in jet energy measurements.
Except for the multijet background, which is model using a multijet enriched data sample, all the other backgrounds are described using
Monte Carlo simulated events.
The multijet-enriched sample used in the D0 analysis is defined by the $NN_{\tau}$ output value: $0.3 \le NN_{\tau} \le 0.7$.
CDF uses events where the identified $\tau$ has exactly two tracks in the $10\,^{\circ}$ signal cone, while $\tau$'s candidate in the signal sample
have either one or three tracks in the signal cone.\\
The multijet background is overwhelming in the sample, if only the presence of a $\tau$, $\met$ and two high $\pt$ calorimeter jets ($\et \ge 20\ \GeV$) is required.
D0 analysis removes most of the it using the $\met$ significance, which takes into account the resolution of the jet energies to assess 
the significance of the observed $\met$ relative to the expected fluctuations in measured jet energies (see  \ref{fig:whtnbb_d0}{\rm a}):
the events must satisfy $\met^{Sign} \ge 4.5$, and $\Delta\Phi (\mpt,\met) \le 0.2 \ rad$, where $\mpt$ is the missing transverse momentum,
calculated using only tracks information.
CDF uses $\met^{Perp}$, the component of the $\met$ that is perpendicular to the closest object (either a $\tau$ or a jet), to reduce the multijet background. Figure 
\ref{fig:whtnbb_cdf}{\rm a} shows the distribution of $\met^{Perp}$ before any b-tagging requirement is applied.\\

\indent Since the low mass  Higgs boson ($m_{H} \le 135 \ \gevcc$) decays mostly into $b\bar{b}$ pairs ( BR($H \to b\bar{b}$) $\sim$ 72\% for $m_{H} =115 \ \gevcc$ ),
requiring that at least one of the jet is consistent with originating from a $b$-quark improves the $S/\sqrt{B}$ by a factor of 2.\\
D0 cuts on the outputs of a Neural Network algorithm to identify $b$-quark jets (for more details,see \cite{D0_btag}). 
Two classes of events are selected: either both the two jets are tagged with a lower efficiency by higher purity (``loose tag''), or only one
of the jet is tagged with a lower efficiency but higher purity (``very tight tag'').\\
CDF definition of a b-tagged jet relies on a secondary vertex tagging algorithm called SecVtx (see \cite{svx}). The algorithm 
reconstructs a secondary vertex inside a jet using tracks with large impact parameter. A jet is $b$-tagged if
the significance of the displacement of the secondary vertex from the primary one is greater than 3. 
Events must contain at least one $b$-tagged jet.
\\
\indent In the low mass Higgs boson searches, the invariant mass of the two jets corresponds to the Higgs boson mass, and it is the most powerful observable
to discriminate the signal from the backgrounds.\\
CDF uses a binned likelihood fit of the di-jet invariant mass distribution to test for a potential Higgs boson signal, and, in its absence, set 
the 95\% C.L. limit. Events are classified in two statistically independent channels: events with exactly one $b$-tagged jet, and
events with two $b$-tagged jets.
The two channels have similar sensitivity,
since the $\met^{perp}$ cut value was optimized for the best sensitivity separately in the two samples.\\
D0 combines the information carried by the di-jet invariant mass with other kinematical observable, such as the energy and pseudorapidity of the individual
jets, of the $\tau$ candidate, and the \met, in a boosted decision tree (BDT), to achieve the best separation between signal and
backgrounds. The BDT is trained for type-I and type-II $\tau$'s, for events with two or three jets and one or two $b$-tagged jets, separately, for a
total of 8 independent channels.\\
Figures \ref{fig:whtnbb_d0}{\rm c} and \ref{fig:whtnbb_cdf}{\rm c} shows the 95\% C.L. upper limit on the Higgs boson production rate time BR($H \to b\bar{b}$) as 
a function of the Higgs boson mass for D0 and CDF respectively. The D0 analysis results with a observed(expected) limit for $m_{H} = 115 \ \gevcc$ 
of 14.1(22.4) time the S.M. expectation. CDF analysis sets an upper limit of 28.7(46.6) for the same mass hypothesis.

\section{Conclusion}
We presented here the most recent results for the Higgs boson searches at the Tevatron, performed in final states that include $\tau$ leptons.
The main challenge, in these analysis, is due to the necessary compromise between retaining as much Higgs boson signal as possible
 and minimizing the contribution of the multijet background, that is difficult to model.\\
Although these channels set limits on the Higgs boson production cross section 10 time greater than the Standard Model expectation,
in the current effort to extend the potential sensitivity at the Tevatron, their contribution is important.

\bigskip 

\begin{thebibliography}{99}   
\bibitem{Higgs}P. W. Higgs, Phys. Lett. 12, 132 (1964).
\bibitem{EWKFit} LEP Electroweak Working Group, {\tt http://lepewwg.web.cern.ch/LEPEWWG/}
\bibitem{lephiggs}    ALEPH, DELPHI, L3, OPAL.  The LEP Working Group for Higgs Boson Searches, Phys. Lett {\bf B565} 61 (2003).  
\bibitem{combination} Tevatron New Phenomena and Higgs Working Group for CDF and D0 collaborations [arXiv:hep-ex/1107.5518v2]
\bibitem{CDF} P. T. Lukens (CDF IIb) (2003), FERMILAB-TM-2198
\bibitem{D0} V.M. Abazov {\rm et al} (D0 collaboration), Nucl. Instrum. Methods in Phys. Res. A {\bf 565},463 (2003)
\bibitem{BDT}  L. Breiman et al., Classication and Regression Trees (Wadsworth, Stamford CT, 1984); D. Bowser-Chao and  D.L. Dzialo, Phys. Rev D 47, 1900 (1993)
\bibitem{htt_cdf} T. Aaltonen {\rm et at} (CDF Collaboration), CDF Confernce Note 10409
\bibitem{tauD0}V.M. Abazov {\rm et al.} (D0 collaboration), Phys. Lett. B {\bf 670}, 292 (2009) 
\bibitem{htt_d0} V. Abazov et al. D0 Conference Note 6171
\bibitem{SVM} 
C. Cortes and V. Vapnik, ``Support vector networks'', Machine Learning, 20, 273 (1995). \\
V. Vapnik, ``The Nature of Statistical Learning Theory'', Springer Verlag, New York, 1995. \\
C.J.C. Burges,''A Tutorial on Support Vector Machines for Pattern Recognition'', Data Mining and
Knowledge Discovery, 2, 1 (1998).
\bibitem{lltt} T. Aaltonen {\rm et at} (CDF Collaboration), CDF Confernce Note 10500
\bibitem{HWW_cdf} T. Aaltonen {\rm et at} (CDF Collaboration), CDF Confernce Note 10597
\bibitem{HWW_d0} V. Abazov et al. (D0 Collaboration) D0 Conference Note 6179
\bibitem{whtnbb_cdf} T. Aaltonen {\rm et at} (CDF Collaboration), CDF Confernce Note 10648
\bibitem{whtnbb_d0} V. Abazov et al. D0 Conference Note 5977
\bibitem{D0_btag} T. Scanlon, ``b-Tagging and the Search for Neutral Supersymmetric Higgs Boson at D0'', FERMILAB-THESIS-2006-43
\bibitem{svx} V. Abazov et al. D0 Conference Note 6171


\end{thebibliography}

\end{document}